%% file: imt.tex
\documentclass[acmtog,nonacm]{acmart}

\usepackage{balance}
\usepackage{booktabs} 
\usepackage{listings}
\usepackage{color}
\usepackage{subcaption}
\usepackage{cleveref}
\usepackage{textcomp}
\usepackage{graphicx}

\usepackage{titlesec}
\usepackage{lipsum}
\usepackage{MnSymbol}

\usepackage[ruled, lined, linesnumbered, commentsnumbered, longend]{algorithm2e}
\usepackage{booktabs,siunitx,array,threeparttable}
\usepackage{paralist}
\usepackage{amsthm}

\definecolor{dkgreen}{rgb}{0,0.6,0}
\definecolor{gray}{rgb}{0.5,0.5,0.5}
\definecolor{mauve}{rgb}{0.58,0,0.82}

\begin{document}

\title{Beyond Pairwise Learning-To-Rank At Airbnb}

\author{Malay Haldar \\ Daochen Zha \\ Huiji Gao \\ Liwei He \\ Sanjeev Katariya}
\makeatletter
\let\@authorsaddresses\@empty
\makeatother




\renewcommand{\shortauthors}{Malay Haldar et al.}

\begin{abstract}
\textbf{Abstract:} There are three fundamental asks from a ranking algorithm: it should {\it scale} to handle a large number of items, sort items {\it accurately} by their utility, and impose a {\it total order} on the items for logical consistency. But here’s the catch---no algorithm can achieve all three at the same time. We call this limitation the SAT theorem for ranking algorithms. Given the dilemma, how can we design a practical system that meets user needs? Our current work at Airbnb provides an answer, with a working solution deployed at scale.

We start with pairwise learning-to-rank (LTR) models---the bedrock of search ranking tech stacks today. They scale linearly with the number of items ranked and perform strongly on metrics like NDCG by learning from pairwise comparisons. They are at a sweet spot of performance vs. cost, making them an ideal choice for several industrial applications. However, they have a drawback---by ignoring interactions between items, they compromise on accuracy. 

To improve accuracy, we create a “true” pairwise LTR model---one that captures interactions between items during pairwise comparisons. But accuracy comes at the expense of scalability and total order, and we discuss strategies to counter these challenges.

Traveling further along the road to greater accuracy, we take each item in the search result, and compare it against the rest of the items along two dimensions: (1) {\it Superiority}: How strongly do searchers prefer the given item over the remaining ones? (2) {\it Similarity}: How similar is the given item to all the other items? This forms the basis of our “all-pairwise” LTR framework, which factors in interactions across all items at once. Looking at  items on the search result page all together---superiority and similarity combined---gives us a deeper understanding of what searchers truly want. We quantify the resulting improvements in searcher experience through offline and online experiments at Airbnb.
\end{abstract}

%
%

\maketitle

\input{imtbody-conf}

\bibliographystyle{ACM-Reference-Format}

\balance 
\bibliography{imt-bibliography}

\end{document}

%% file: imtbody-conf.tex
\section{Introduction}
Search ranking orders a corpus of items to maximize some combination of utilities for the searcher. It’s the tool we rely on daily to cut through endless choices—whether it’s finding some information, a meal, a job, a vacation spot, or a place to stay. At the heart of search ranking is the learning-to-rank (LTR) algorithm, which learns from past searcher decisions to improve future results. A straightforward way to learn from past examples is to frame it as a binary classification task, where items that satisfied the searchers' needs are treated as positive, the remaining choices that were presented to the searcher are labeled negative, and a model is trained to distinguish between the two. However, many industrial scale applications use not the simple binary classification, but a variant of the formulation known as pairwise LTR.

The seminal paper that introduced the Bradley-Terry-Luce model and analyzed pairwise LTR appeared in 1957 ~\cite{btl}, with neural network applications to the problem emerging in a 1994 paper ~\cite{siamese}. Since then, pairwise LTR has found adoption at major internet companies ~\cite{sculley2009large} \cite{Senthilkumar2024} \cite{listloss} \cite{pairlinkedin}, including Airbnb ~\cite{kdd19}. What has kept this decades-old area of research active and thriving? It's the increasing availability of data, advances in computational power, and the growing business value of improved rankings. This paper presents one step forward in this long narrative arc, highlighting recent advancements in ranking at Airbnb that go beyond traditional pairwise LTR.

We start by examining the properties that make pairwise LTR scalable and effective. The formalism introduced in Section~\ref{sec:pltr} is a stepping stone for the sections ahead, where we address the limitations of pairwise LTR that impact its accuracy. We make improvements in two steps, first considering interactions between two listings at a time in Section~\ref{sec:truepair}, then in Section~\ref{sec:allpair} scaling up to interactions across all listings. Section~\ref{sec:multiobj} extends the framework from optimizing for a single objective to optimizing for multiple objectives. We cite related work embedded in context of the discussion throughout the paper. The final section covers experiments performed at Airbnb to compare the frameworks developed, and their outcomes. But first, the SAT theorem.

\section{SAT Theorem for Ranking Algorithms} \label{sec:sat}
A practical ranking algorithm needs to meet three requirements:
\begin{itemize}
\item {\it Scalability}: From a systems perspective, ranking must be able to handle thousands to millions of items---typically within a latency budget of $O(100)$ milliseconds. 
\item {\it Accuracy}: From a user perspective, items with higher utility should rank higher. This ensures the most useful items appear in positions that get the most attention. Accuracy is evaluated using ranking metrics like NDCG and MAP.
\item {\it Total order}: To make ranking mathematically well-defined, it needs to create a total order~\cite{wiki:totalordering}. This means for any two items, $a$ and $b$:
\begin{enumerate}
\item $a \leq a$ (reflexive).
\item If $a \leq b$ and $b \leq c$ then $a \leq c$ (transitive).
\item If $a \leq b$ and $b \leq a$ then $a = b$ (antisymmetric).
\item $a \leq b$ or $b \leq a $ (strongly connected).
\end{enumerate}
\end{itemize}

In the spirit of the CAP theorem~\cite{wiki:cap} for distributed systems, the SAT theorem states that a ranking algorithm can only fulfill two of these three requirements at a time. Tversky's insights from choice theory~\cite{tversky1972elimination} point to the root cause for this tension---item utility depends on other items, which is hard to model while guaranteeing total order and scalability.

We start by describing how pairwise LTR deals with the SAT theorem.

\section{Pairwise LTR: A Brief Overview} \label{sec:pltr}
Pairwise LTR is at the core of search ranking at Airbnb. Users search by entering a location, check-in/out dates, and the number of guests. In response, the ranking algorithm organizes the available listings based on several factors. How this ranking algorithm evolved over the years is captured in ~\cite{kdd19}, ~\cite{kdd20}, ~\cite{kdd23}, ~\cite{cikmdiversity}, and ~\cite{kdd24}.

The base objective of the pairwise LTR model is to order the listings by their probability of getting booked. The model is trained on pairs of booked and not-booked listings extracted from logged search results following this setup:\\
$l_{\text{booked}}$ : features of the booked listing\\
$l_{\text{notBooked}}$ : features of the listing not booked\\
$f(l)$ : an MLP (multi-layer perceptron) with listing $l$ as input.\\ 
$\theta(x) = (1 + e^{-x})^{-1}$, the sigmoid function \\

The training process minimizes the loss:
\begin{equation}\label{eq:pairloss}
\text{loss} = -\log(\theta(f(l_{\text{booked}}) - f(l_{\text{notBooked}})))
\end{equation}
which is the binary cross-entropy loss with label set to 1. Throughout the paper, MLPs like $f(\cdot)$ also take user and query features as input, but we’ve skipped them for readability. After training, the model produces two quantities:

\begin{equation}\label{eq:pairlogit}
\text{Pairwise logit, } \text{logit}(l_a, l_b) = f(l_a) - f(l_b) 
\end{equation}
\begin{equation}\label{eq:pairprob}
\text{Pairwise probability, } P(l_a > l_b) = \theta(\text{logit}(l_a, l_b))
\end{equation}

For ranking $N$ listings, pairwise LTR assigns each listing $l_i$ a score using $f(l_i)$ and then sorts listings by their scores. Since $f(\cdot)$ is {\it univariate}, scoring one listing at a time, pairwise LTR scales linearly. The MLP $f(\cdot)$ provides a deterministic mapping from listing features to a real valued score. The total order of listings can be obtained from the total order of the corresponding scores.

The place where pairwise LTR misses out is accuracy. Here's why. Ordering the listings by $f(\cdot)$ is equivalent to ordering them by the pairwise booking probabilities. The Bradley-Terry equation ~\cite{wiki:bradleyterry} links these pairwise booking probabilities to the individual (pointwise) booking probabilities $P(l_a)$ and $P(l_b)$:
\begin{equation}
P(l_a > l_b) = P(l_a) / (P(l_a) + P(l_b))
\end{equation}
Hence,  $f(l_a) > f(l_b)$ implies $P(l_a) > P(l_b)$, and pairwise LTR ends up ordering listings by their pointwise booking probabilities. The key limitation here is that these probabilities are treated as if they depend only on the listing itself. In reality, the booking probability of a listing is influenced by the entire set of $N$ listings shown to the user~\cite{tversky1972elimination}.

For a more precise understanding of pairwise LTR's limitations, let's focus on the properties of the pairwise logit described in Equation~\ref{eq:pairlogit}. If we take three listings, $l_a$, $l_b$, and $l_c$, it's easy to show that:
 \begin{equation} 
\text{logit}(l_a, l_b) + \text{logit}(l_b, l_c) = \text{logit}(l_a, l_c)
 \end{equation}
 This proves that for a set of $N$ listings, all the $N(N-1)/2$ pairwise logits align along a single straight line. This characteristic is known as collinearity. Collinearity does {\it not} mean that pairwise LTR is learning a linear relationship between the input features of the listings. Those relationships are actually managed by $f(\cdot)$, a non-linear MLP. Collinearity simply indicates that the pairwise-logit space---created by comparing all listing pairs---is limited to one dimension.

This restriction on the logit space prevents pairwise LTR from accurately modeling booking probabilities for users. Let’s break it down with an example. Imagine two listings, $l_x$ and $l_y$, that are identical apartments in the same building, but $l_x$ is \$10 cheaper. Naturally, most searchers would prefer $l_x$, so $\text{logit}(l_x, l_y)$ would have a high positive value, i.e., $\text{logit}(l_x, l_y) \gg 0$.

\begin{figure}
\includegraphics[height=0.75in, width=1.773in]{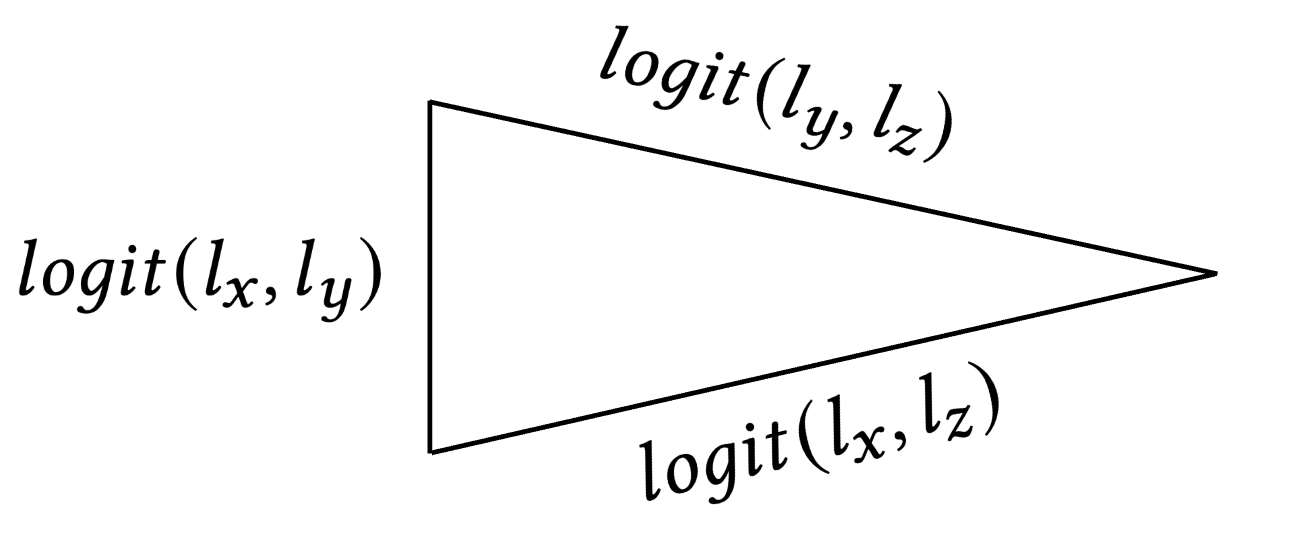}
\caption{\textmd{The relation $\text{logit}(l_x, l_y) \gg 0 \land \text{logit}(l_x, l_z) \approx \text{logit}(l_y, l_z)$ cannot be captured in a 1D space. It is represented by the triangle inequality $\text{logit}(l_x, l_y) + \text{logit}(l_y, l_z) > \text{logit}(l_x, l_z)$.}}
\label{fig:collinearity}
\end{figure}

Now add a third listing, $l_z$, that’s quite different in terms of location, price, and amenities. The values of $\text{logit}(l_x, l_z)$ and $\text{logit}(l_y, l_z)$ would be similar because users who favor $l_x$ over $l_z$ would also favor $l_y$ over $l_z$. This means $\text{logit}(l_x, l_z) \approx \text{logit}(l_y, l_z)$. If we apply collinearity, we’d get
\begin{equation*}
\text{logit}(l_x, l_y) + \text{logit}(l_y, l_z) \approx \text{logit}(l_y, l_z)
\end{equation*} 
which implies $\text{logit}(l_x, l_y) \approx 0$. This contradicts the earlier result of $\text{logit}(l_x, l_y) \gg 0$, proving that collinearity doesn’t apply in this case. Figure~\ref{fig:collinearity} illustrates why pairwise comparisons can’t always fit into a 1D logit space.

\section{True-Pairwise LTR} \label{sec:truepair}
In order to pursue accuracy, we break free from the restrictive 1D logit space of pairwise LTR and build a “true” pairwise model $g(l_a, l_b)$. Unlike the {\it univariate} pairwise LTR model $f(\cdot)$, the true-pairwise model $g(\cdot)$ is {\it bivariate}. It takes both the listings being compared as inputs, allowing it to learn interactions between the two listings. When training this model on an example pair of booked and not-booked listings $\{l_{bk}, l_{nbk}\}$, the loss is computed as:
\begin{equation} \label{eq:tploss}
\text{loss} = -\log(\theta(g(l_{bk}, l_{nbk}))) - \log(1 - \theta(g(l_{nbk}, l_{bk})))
\end{equation}
which is the symmetric binary cross-entropy loss for predicting that $l_{bk}$ is preferred over $l_{nbk}$.

The pairwise logit is directly implemented by the true-pairwise LTR model as:
\begin{equation}
\text{logit}(l_a, l_b) =g(l_a, l_b)
\end{equation}
Because $g(\cdot)$ is a general bivariate function, it can represent the types of relationships shown in Figure~\ref{fig:collinearity}. As an MLP, $g(\cdot)$ is flexible enough to learn any interaction between $l_a$ and $l_b$. This improves the accuracy of the pairwise booking probability $\theta(g(l_a, l_b))$. However, as stated by the SAT theorem, this increase in accuracy comes at a cost: reduced scalability and the loss of total order. We dive into strategies to restore total order and mitigate the loss of scalability next.

\subsection{Total Order}
\begin{figure*}
\begin{subfigure}[t]{0.45\textwidth}
\centering
\includegraphics[height=1.8in, width=1.8in]{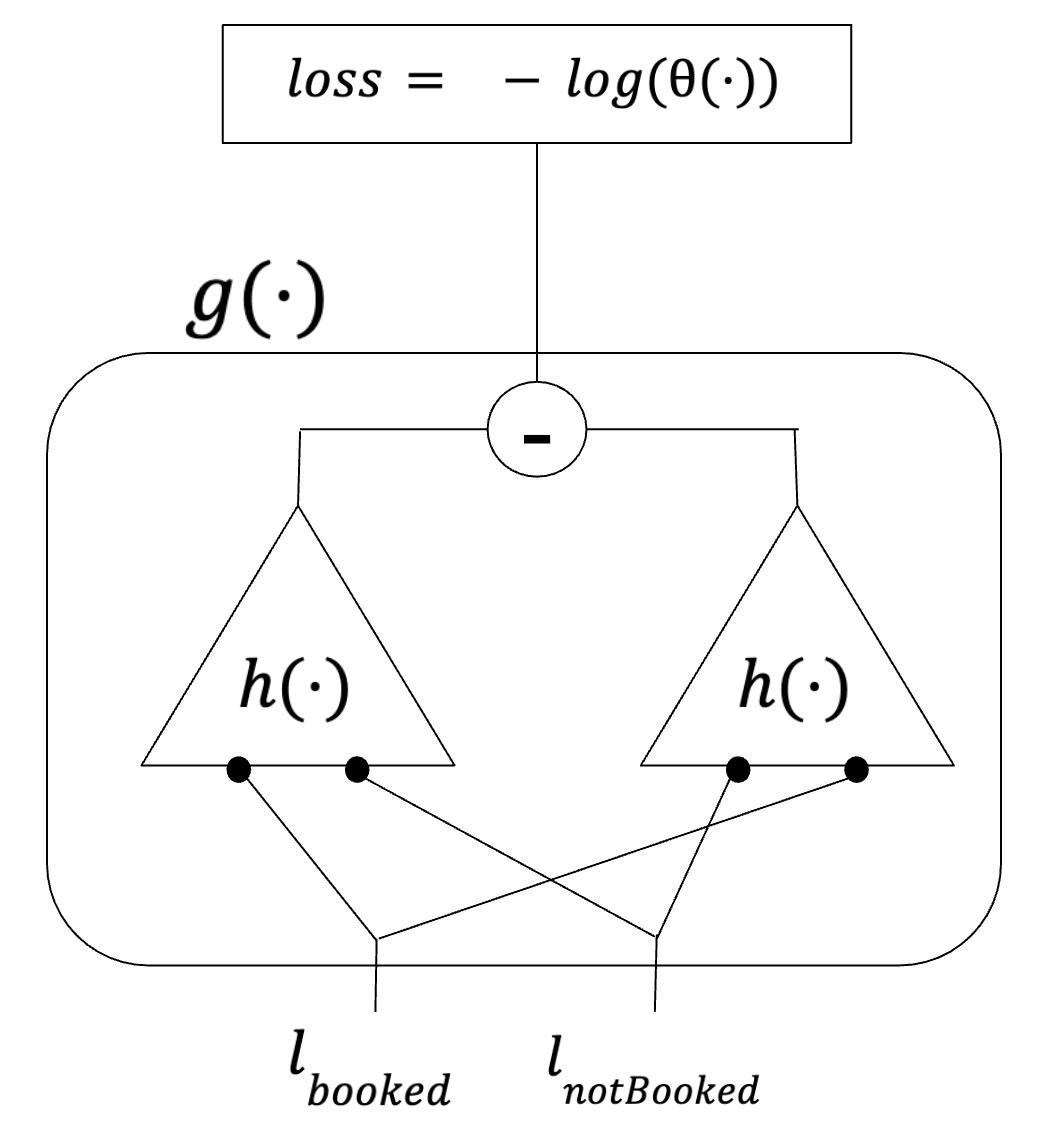}
\caption{\textmd{True-pairwise training, optimizing the loss from Equation~\ref{eq:tploss}.}}
\label{fig:truepairtrain}
\end{subfigure}
\hfill
\begin{subfigure}[t]{0.45\textwidth}
\centering
\includegraphics[height=2.3in, width=2.3in]{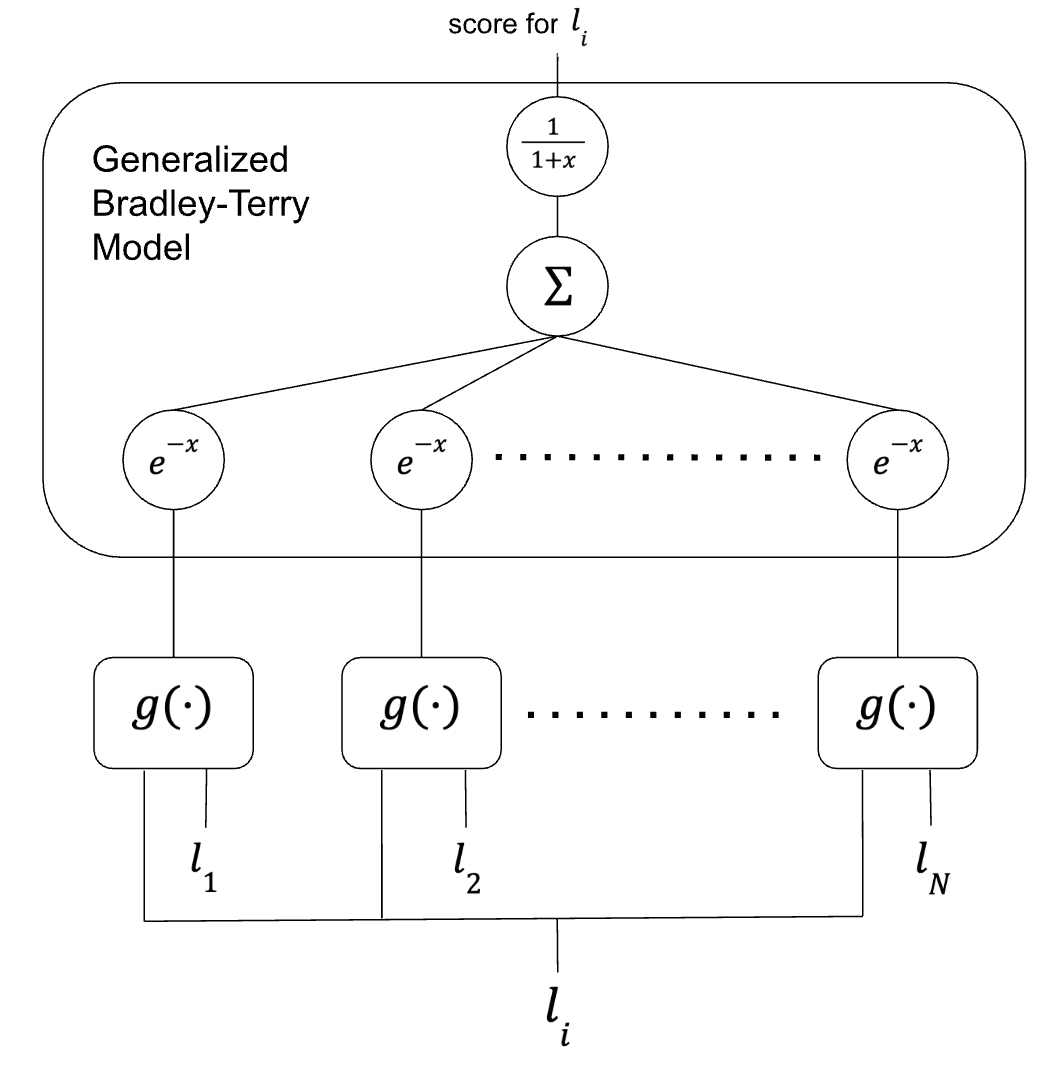}
\caption{\textmd{True-pairwise scoring utilizing Equation~\ref{eq:genbradter}.}}
\label{fig:truepairscoring}
\end{subfigure}
\caption{True-pairwise LTR.}
\end{figure*}

The true-pairwise LTR model doesn’t naturally enforce the conditions needed for total order (described in Section~\ref{sec:sat}). To tackle this, we make $g(\cdot)$ anti-commutative, meaning: $g(l_a, l_b) = -g(l_b, l_a)$. Enforcing anti-commutativity is a practical way to guarantee three of the four conditions—reflexivity, antisymmetry, and strong connectivity—as directly enforcing them is hard. To make $g(\cdot)$ anti-commutative, we define it as:
\begin{equation}\label{eq:tpbreak}
g(l_a, l_b) = h(l_a, l_b) - h(l_b, l_a)
\end{equation}
Here, $h(\cdot)$ is an MLP that learns interactions between the two listings. This approach ensures $g(l_a, l_b) = -g(l_b, l_a)$ while keeping the flexibility to capture complex relationships between $l_a$ and $l_b$. Figure~\ref{fig:truepairtrain} illustrates how the true-pairwise LTR model is trained.

Even with anti-commutativity in place, we still need to enforce transitivity to fully establish total order. Without transitivity, $g(l_a, l_b)$ cannot be directly plugged into popular sorting algorithms like Quicksort or Mergesort.

To solve this problem given $N$ listings, for each listing $l_i$, we collect the $N-1$ scores $g(l_i, l_1), g(l_i, l_2), \dots , g(l_i, l_n)$. Given that we have established anti-commutativity for $g(\cdot)$, we can apply a simpler version of the generalized Bradley-Terry model (Equation 10 from ~\cite{general_bradley_terry}):
\begin{equation}\label{eq:genbradter}
\text{score}(l_i) = \frac{1} { 1 + \sum _{j \neq i, j = 1}^{N}e^{-g(l_i,l_j)} }
\end{equation}

These scores assign each listing a real value and are invariant to how the listings are ordered initially. Since real-valued scores naturally fulfill the requirements for total order, they can now be used with standard sorting algorithms like Quicksort or Mergesort. Figure~\ref{fig:truepairscoring} illustrates how a listing is scored in the true-pairwise LTR framework.

Our approach can be contrasted with the true-pairwise LTR method proposed in ~\cite{nardini27learning}, where the $N-1$ scores are averaged for a listing. By leveraging the generalized Bradley-Terry model, we achieve a higher accuracy of booking probability due to the performance bounds provided in ~\cite{general_bradley_terry}. The two approaches are compared under experimental results.

\subsection{Scalability}
\begin{figure}
\includegraphics[height=1.7in, width=3.4in]{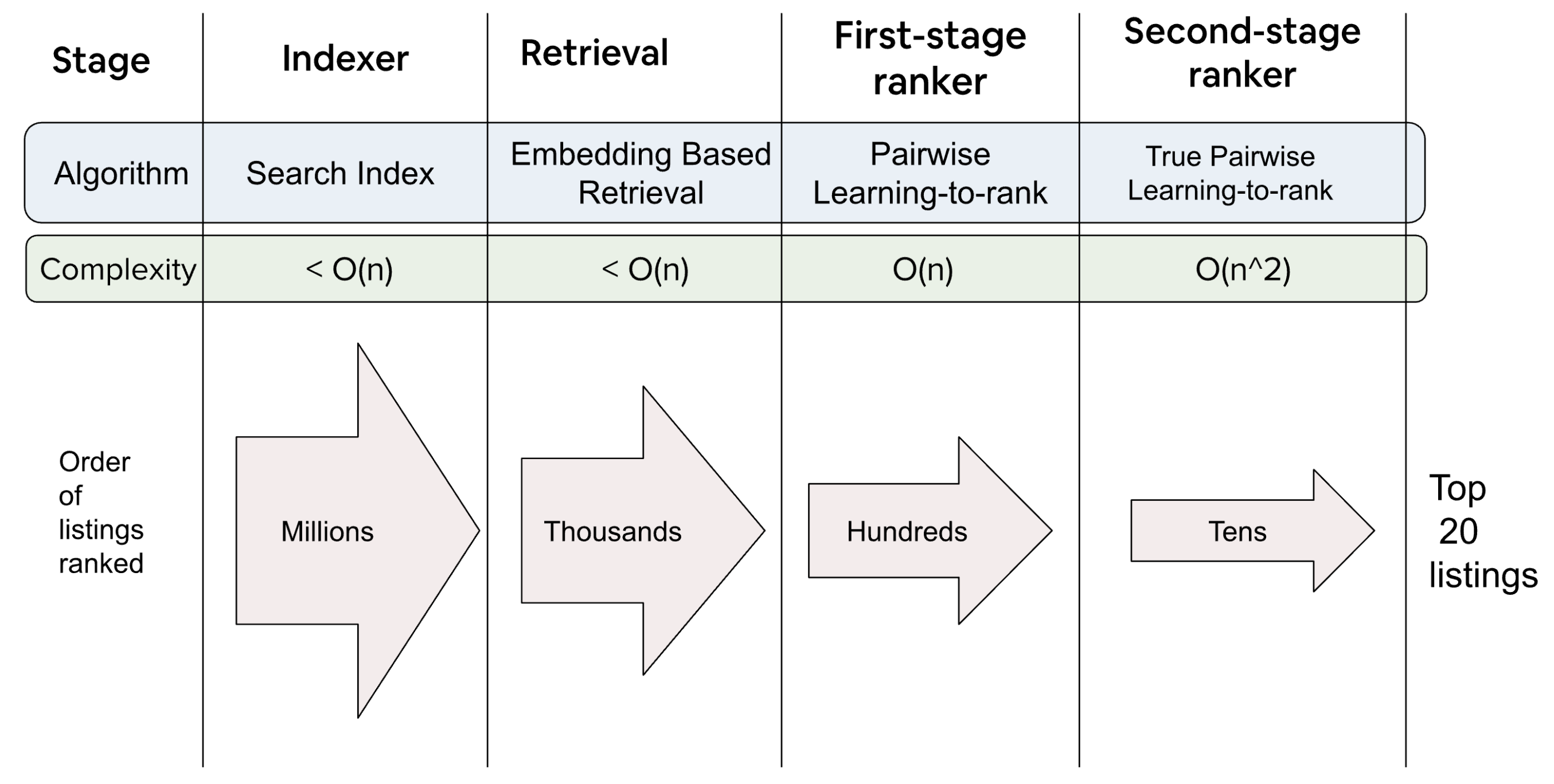}
\caption{\textmd{Anatomy of a multi-stage ranker.}}
\label{fig:rankerstages}
\end{figure}

With $N$ listings, true-pairwise LTR requires $N(N-1)/2$ unique evaluations of $g(\cdot)$, one for each pair of listings. In contrast, pairwise LTR only needs $N$ evaluations of $f(\cdot)$. Because of its quadratic complexity, we use true-pairwise LTR only in the final stage of a multi-stage ranking system, after the number of listings has been reduced to $O(10)$. This leads to a hybrid system where the first-stage ranker is still owned by pairwise LTR. Figure~\ref{fig:rankerstages} outlines the complexity levels across different ranking stages. For more on algorithms suited to the final ranking stage, see ~\cite{ijcai2022p771}.

\section{All-Pairwise LTR} \label{sec:allpair}
Bivariate true-pairwise LTR improves the accuracy of univariate pairwise LTR by accounting for interactions between the two listings being compared. Building on this idea, we introduce a {\it multivariate} LTR framework that considers how each listing interacts with all the other listings at once. This approach better aligns with the user behavior described in choice theory~\cite{tversky1972elimination} and leads to improved accuracy of booking probability predictions.

Scoring a listing in true-pairwise LTR (Figure~\ref{fig:truepairscoring}) consists of two steps:
\begin{enumerate}
\item Calculating the $N-1$ pairwise logits using the true-pairwise LTR model $g(\cdot)$.
\item Combining these pairwise logits into one final score using the generalized Bradley-Terry model (Equation~\ref{eq:genbradter}).
\end{enumerate}
In all-pairwise LTR, we replace both these steps by a single model. This model not only takes the listing being scored as input, but the remaining $N-1$ listings as well to compute interactions.

The generalized Bradley-Terry model, used in the second step of true-pairwise LTR, lacks adaptability since it doesn’t have learnable parameters. To address this, we replace it with an MLP named the all-pairwise logit network (APLN). During training the APLN generates logits, which we pair for booked and non-booked listings, and then optimize the loss in Equation~\ref{eq:pairloss}.

Next we turn our attention to the features that should feed the APLN. To efficiently capture how a listing interacts with the other $N-1$ listings, we create two types of features:
\begin{itemize}
\item {\it Superiority features}: Due to its $O(N^2)$ complexity, the all-pairwise LTR model serves as the second-stage ranker, reordering the top $N$. We still have the pairwise LTR model as the first-stage ranker, and therefore the pairwise LTR logits available for each listing. We reuse these logits from the first-stage ranker as features for the second-stage ranker. We take pairwise differences of the logits and apply a sigmoid, generating a feature in the $(0,1)$ range. This in effect captures the relative superiority of the current listing compared to the rest. 
\item {\it Similarity features}: The superiority features fail to capture a crucial piece of information---the similarity between listings. Note the violation of collinearity discussed in Section~\ref{sec:truepair} relied on similar listings. Accounting for listing similarities is essential to diversify search results, discussed in our previous work ~\cite{cikmdiversity}. To create similarity features, we first map the listing feature vectors to embeddings. Dot product of the embeddings measure similarity, and a softmax normalizes the dot product to the $(0, 1)$ range. 
In our earlier work ~\cite{cikmdiversity}, we manually built a scoring function on top of these similarities. In all-pairwise LTR, we let the APLN learn this scoring function.
\end{itemize}

But before we can feed the superiority-similarity features into the APLN, we need to overcome one more obstacle. Recall that to implement a valid comparator, the true-pairwise LTR model needed to be anti-commutative. For all-pairwise LTR where we input multiple listings, this requirement generalizes to ensuring input permutation invariance.

If superiority-similarity features are directly fed into the APLN, input permutation invariance cannot be guaranteed. To see this, consider a single neuron of the input layer of the APLN. Let the $i$th input (e.g. superiority over $l_i$) be $x_i$, the $i$th weight be $w_i$, and $b$ be the layer bias. The neuron’s output, $x_i*w_i+b$, changes if $l_i$ is reordered. To make the neuron’s output invariant to input order, we make the $i$th weight $w_i$ depend on $l_i$. This dependency is modeled through a function $\phi(l_i)$, implemented as an MLP. The modified neuron output, written as $x_i*\phi(l_i)+b$, is now permutation invariant.

\begin{figure*}
\centering
\begin{subfigure}[t]{0.45\textwidth}
\centering
\includegraphics[height=2.3in, width=3.105in]{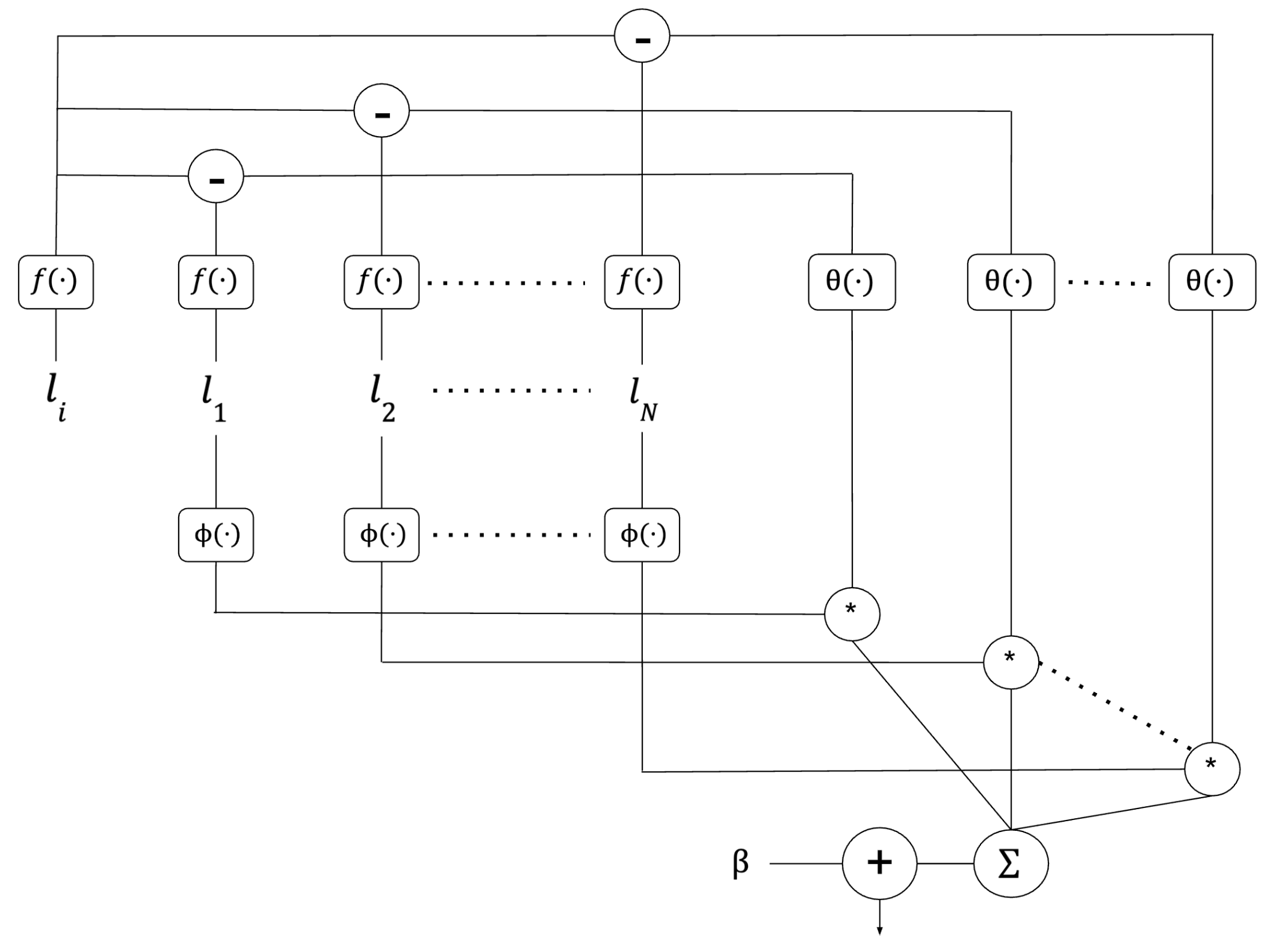}
\caption{\textmd{Feature computing the superiority of the given listing $l_i$ over the remainder of listings $l_1, l_2, \dots l_N$. $f(\cdot)$ is the pairwise LTR first-stage ranker represented by Equation~\ref{eq:pairlogit}. $\theta(\cdot)$ is the sigmoid function. $\phi(\cdot)$ is an MLP which takes the listing features as input. $\beta$ is the bias term.}}
\label{fig:superiority}
\end{subfigure}
\hfill
\begin{subfigure}[t]{0.45\textwidth}
\centering
\includegraphics[height=2.3in, width=3.105in]{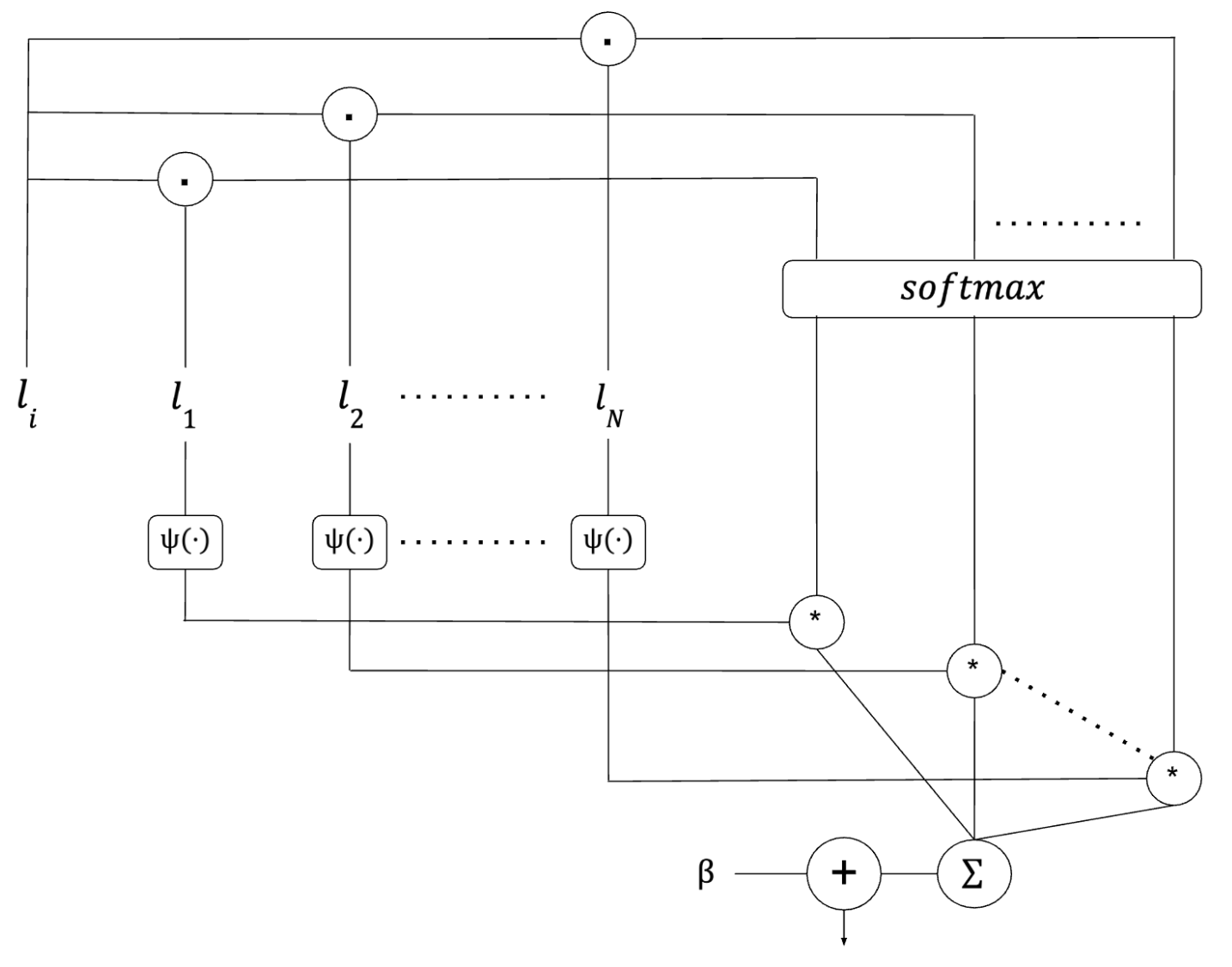}
\caption{\textmd{Feature computing similarity of the given listing $l_i$ to the remainder of listings $l_1, l_2, \dots l_N$ by taking dot products. $\psi(\cdot)$ is an MLP which takes the listing features as input. $\beta$ is the bias term.}}
\label{fig:similarity}
\end{subfigure}
\caption{Superiority and similarity features of APFN.}
\end{figure*}

\begin{figure}
\includegraphics[height=2.3in, width=3.105in]{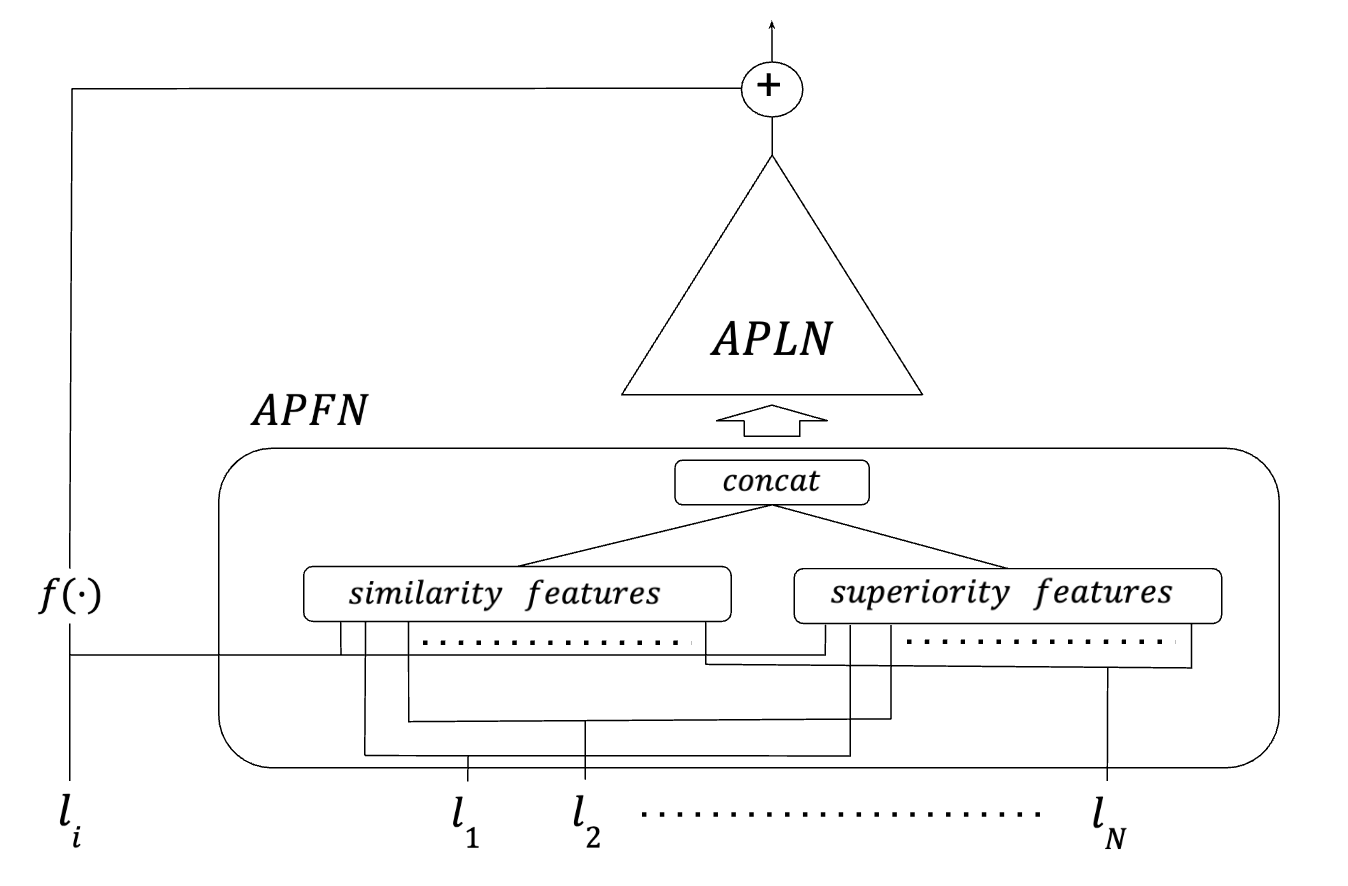}
\caption{\textmd{The all-pairwise architecture. {\it Superiority features} refers to Figure~\ref{fig:superiority}. {\it Similarity features} refers to Figure~\ref{fig:similarity}. Together they form the all-pairwise feature network (APFN). The APFN feeds into the all-pairwise logit network (APLN), an MLP. To produce the final logit, APLN acts as a residual to the logit produced by the pairwise LTR first-stage ranker $f(\cdot)$.}}
\label{fig:allpair}
\end{figure}

An alternative way to achieve permutation invariance, proposed in ~\cite{groupwise}, involves generating and aggregating permutations of item subsets. In our approach, we avoid the computational complexity of creating permutations. The superiority-similarity features, along with the layer imposing input permutation invariance, make up the all-pairwise feature network (APFN). Construction of superiority features is outlined in Figure~\ref{fig:superiority}, while Figure~\ref{fig:similarity} illustrates the similarity features. The overall all-pairwise LTR architecture is shown in Figure~\ref{fig:allpair}.

We designed APFN to learn how the scored listing interacts with the other $N-1$ listings. An alternative to APFN is to use an attention network to construct the features, as explored in ~\cite{googleattn}, ~\cite{rlrerank}, and ~\cite{Pang2020SetRankLA}. This gives us an off-the-shelf solution which is easier to get started with. In the experimental results section to follow, we compare how these methods stack up against each other.

Interactions between the listings scored can also be modeled using RNNs, as shown in ~\cite{rnnrerank} and ~\cite{seq2slate}. Such methods, however, are not input permutation invariant and cannot guarantee total order. Their learning is dependent on the input sequences seen during training and they don't learn a generalized sorting capability.

Another interesting point of comparison for all-pairwise LTR is with neural networks that implement differentiable sorting functions, such as ~\cite{grover2018stochastic} and ~\cite{softsort}. Those too are $O(N^2)$ complexity solutions capable of ranking listings, but their focus is to replace traditional sorting algorithms like Quicksort with a neural network. In contrast, the goal of all-pairwise LTR is to improve ranking by learning interactions between listings, while still using Quicksort or Mergesort to perform the final sorting.

A frequent stumbling block is the apparent resemblance between all-pairwise LTR and its distant cousin, listwise LTR~\cite{listloss}, a univariate model that optimizes the softmax cross-entropy loss. Since listwise LTR is univariate, it cannot account for interactions between listings. Similarly, methods that directly optimize NDCG ~\cite{qin2010general} ~\cite{jagerman2022optimizing} also rely on univariate models, leaving them unable to model listing-to-listing interactions.

All-pairwise LTR maps each listing to a real valued score in a way that doesn’t depend on the order of inputs; these scores subsequently provide a total order. To address scalability, we limit all-pairwise LTR to the second-stage ranking and use it alongside a pairwise LTR model for the first stage. While none of the individual models in this hybrid system can escape the SAT theorem, the system as a whole reaches a high degree of accuracy while maintaining total order and scalability.

\section{Optimizing Multiple Objectives} \label{sec:multiobj}
So far, we've focused on ranking the listings by their booking probability. For ranking systems serving internet-scale applications, conversion is typically the principal, but still one of many objectives. Airbnb is no exception and ranking aims to minimize a range of negative outcomes, while trying to maximize positive outcomes for a trip. As an example to explore multiple objectives, we use trip rating---the 5-star reviews guests leave after their stay.

Determining the relative weights for objectives like booking and trip rating is an extensive topic, and beyond the scope of this paper. For our purposes, we will assume that each objective is given a weight, and the goal of ranking is to maximize the weighted sum of the objectives.

A popular solution to the problem is multi-task learning \cite{slateaware} \cite{kdd23}, where a model predicts multiple objectives through different heads and combines them using weights. This approach makes the model more complex, since it has to handle multiple labels and balance dense and sparse ones. Particularly in our case, true-pairwise and all-pairwise LTR fit only one of the objectives, the booking probability, as searchers naturally compare various listings in search results. These methods aren’t suitable for predicting trip ratings, requiring a separate model to handle that task.

Using separate models for each objective creates challenges---like conflicting outputs and missing interactions between objectives.

We discard the multi-task and multi-model approaches to solve the multiple objectives puzzle. Instead, we unify the objectives by adjusting the training loss. For each training example consisting of a pair of listings $l_{\text{booked}}$ and $l_{\text{notBooked}}$, we compute a weight:
\begin{equation}
\omega_{\text{total}} = 1 + \omega_{\text{tripQuality}}(l_{\text{booked}})
\end{equation}
where a booking gets a base weight of $1$, and trip quality adds to it depending on the reviews of $l_{booked}$. The loss is calculated as:
\begin{equation} \label{eq:weightedloss}
\text{loss} = -\log(\theta(f(l_{\text{booked}}) - f(l_{\text{notBooked}}))) * \omega_{\text{total}}
\end{equation}

Intuitively, instead of treating all pairwise wins equally, we value each win by its trip quality. This gives us a simple and effective way to target multiple objectives within true-pairwise and all-pairwise LTR. We demonstrate the effectiveness of this approach in the experimental results section.

\section{Stability Of Results} \label{sec:stability}
Optimizing ranking for an internet-scale product isn't just about improving metrics like bookings or trip ratings; it also involves addressing qualitative factors. One important consideration is ensuring the stability of search results. In all-pairwise LTR, a listing's fate is intricately tied to its companions in the lineup; any change in the input set can cause the listing scores to change. The input set of listings may change because of innocuous reasons---like small map movements or simple query date refinements. If such minor changes cause noticeable shifts in search results, it creates a jarring user experience.

To enhance the stability of search results, we leverage the pairwise LTR logits generated by the first-stage ranker. Since this ranker is a univariate function that evaluates one listing at a time, its logits are inherently stable. For the second-stage ranker, we treat the all-pairwise LTR model as a residual adjustment layered on top of the first-stage pairwise LTR model (refer to Figure~\ref{fig:allpair}). Adding the multivariate logit as a residual to the stable univariate logit significantly reduces the instability of search results. The experimental results section highlights the improvements in stability achieved using this approach.

\section{Experimental Results}
For our experimental results, we treat pairwise LTR as the baseline and compare it against true-pairwise and all-pairwise LTR. For true-pairwise LTR, we evaluate two versions:
\begin{itemize}
\item {\it true-pairwise-avg}: The final score for a listing is the average of the $N-1$ scores from true pairwise comparisons.
\item {\it true-pairwise-gbt}: The $N-1$ true pairwise comparisons are aggregated using the generalized Bradley-Terry model (Equation~\ref{eq:genbradter}).
\end{itemize}
 For all-pairwise LTR, we evaluate two versions as well:
 \begin{itemize}
 \item {\it all-pairwise-apfn}: This uses the architecture in Figure~\ref{fig:allpair}, where the all-pairwise logit network is fed with superiority-similarity features from the APFN.
 \item {\it all-pairwise-attn}: This substitutes the APFN with an attention network.
 \end{itemize}
All the models are trained on the same training data, which includes pairs of booked and not-booked listings extracted from search logs spanning a year. 

\begin{figure}
\includegraphics[height=1.7in, width=3.4in]{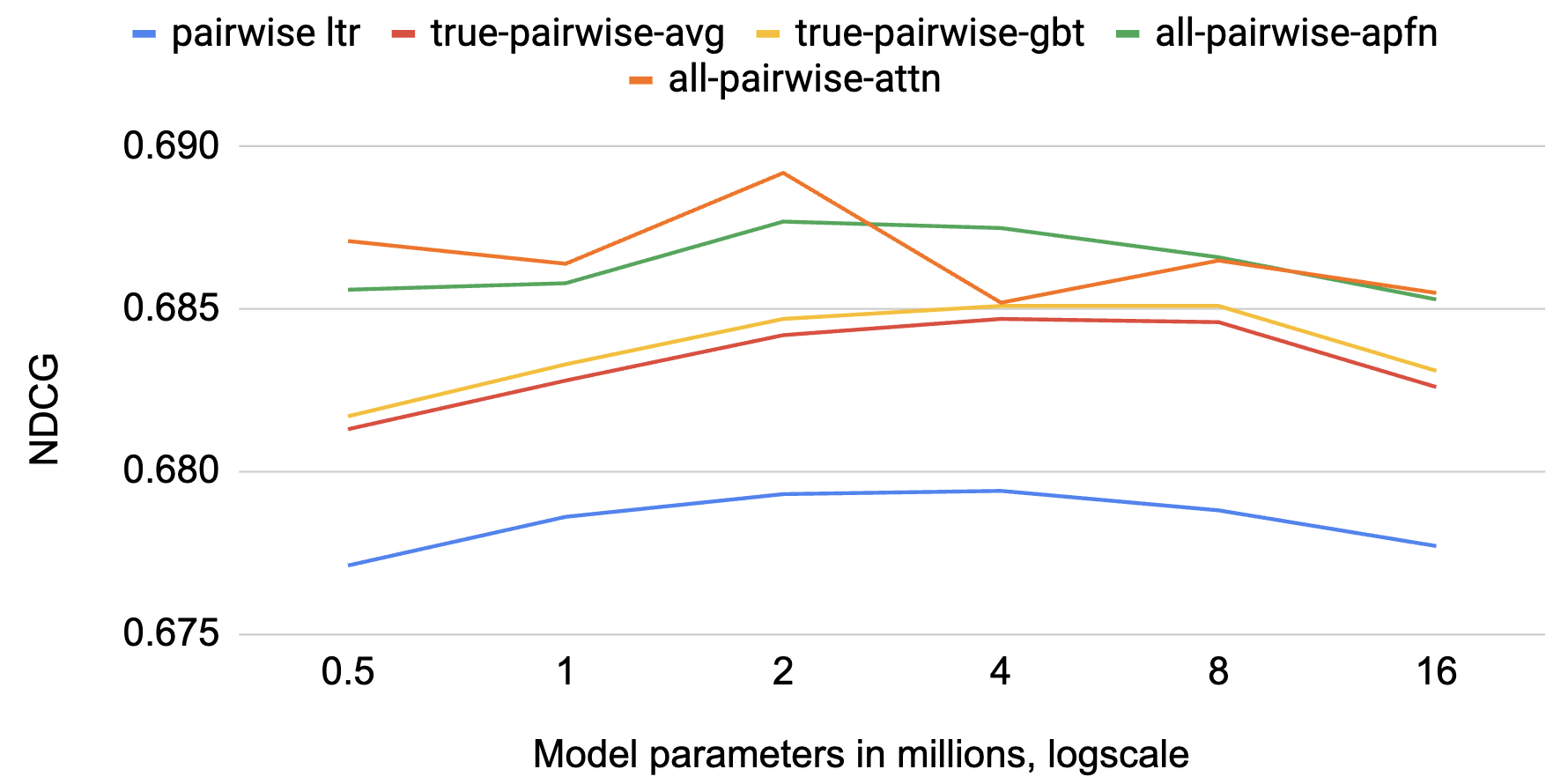}
\caption{\textmd{X-axis: Number of model parameters. Y-axis: Model NDCG.}}
\label{fig:modelparamsndcg}
\end{figure}

We start with offline experiments, using NDCG as the evaluation metric. Booked listings are assigned a relevance of 1, while not-booked listings are assigned a relevance of 0. Given the success of scaling laws for neural language models \cite{kaplan2020scaling} ~\cite{hoffmann2022training}, a frequently-asked question is whether such scaling is applicable to our ranking models as well. Instead of investigating different LTR frameworks, can ranking performance be improved by simply scaling the model parameters of the pairwise LTR model? In our first offline experiment, we study the effect of scaling the number of trainable model parameters (Figure ~\ref{fig:modelparamsndcg}).

The results show that performance improvements from LTR frameworks that better represent searcher behavior cannot be achieved by simple scaling laws. All-pairwise LTR outperforms true-pairwise LTR, which in turn performs better than pairwise LTR, confirming the increase in accuracy predicted by theory. This creates a clear hierarchy: the more effectively a model captures interactions between listings, the better its ranking performance. Additionally, scaling beyond $2$ to $4$ million parameters causes overfitting across all variants, showing that indiscriminate scaling can harm performance.

\begin{figure}
\includegraphics[height=1.7in, width=3.4in]{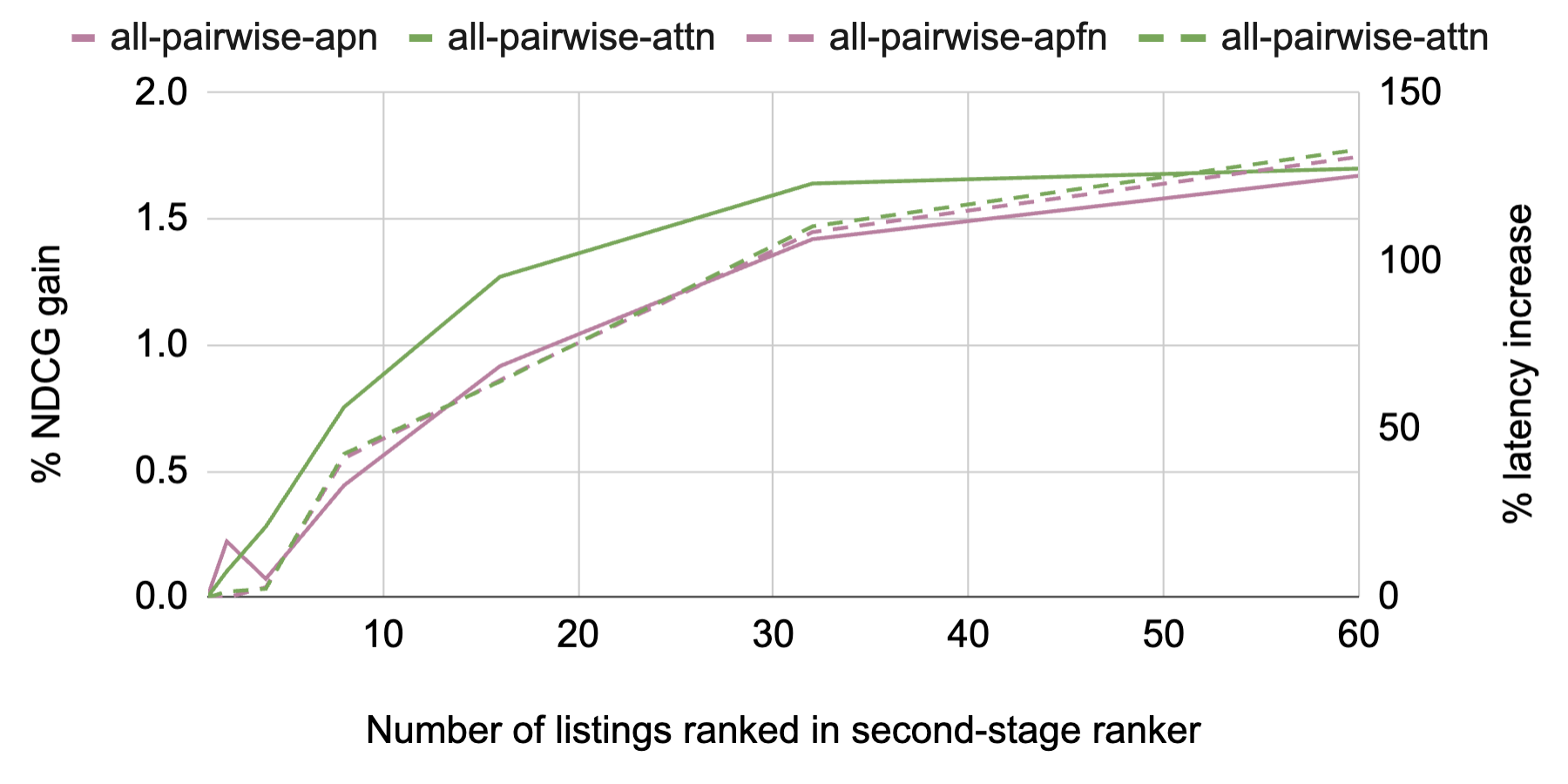}
\caption{\textmd{X-axis: Number of listings reranked. Y-axis left: \% change in NDCG with reranking one listing as baseline. Y-axis right: \% change in latency.}}
\label{fig:numreranked}
\end{figure}

Our second offline experiment focuses on scalability, where we examine the trade-off between cost and benefit for all-pairwise LTR. Figure ~\ref{fig:numreranked} plots the relationship between the number of listings reranked by all-pairwise LTR in the second-stage ranker, and the resulting change in NDCG and latency. The growth in NDCG is sub-linear and flattens beyond a certain point, but the growth in latency continues to rise dramatically. Consequently, there is an optimal trade-off between benefit vs. cost, and we settled at reranking $60$ listings in the second-stage ranker.

\begin{figure}
\includegraphics[height=1.7in, width=3.4in]{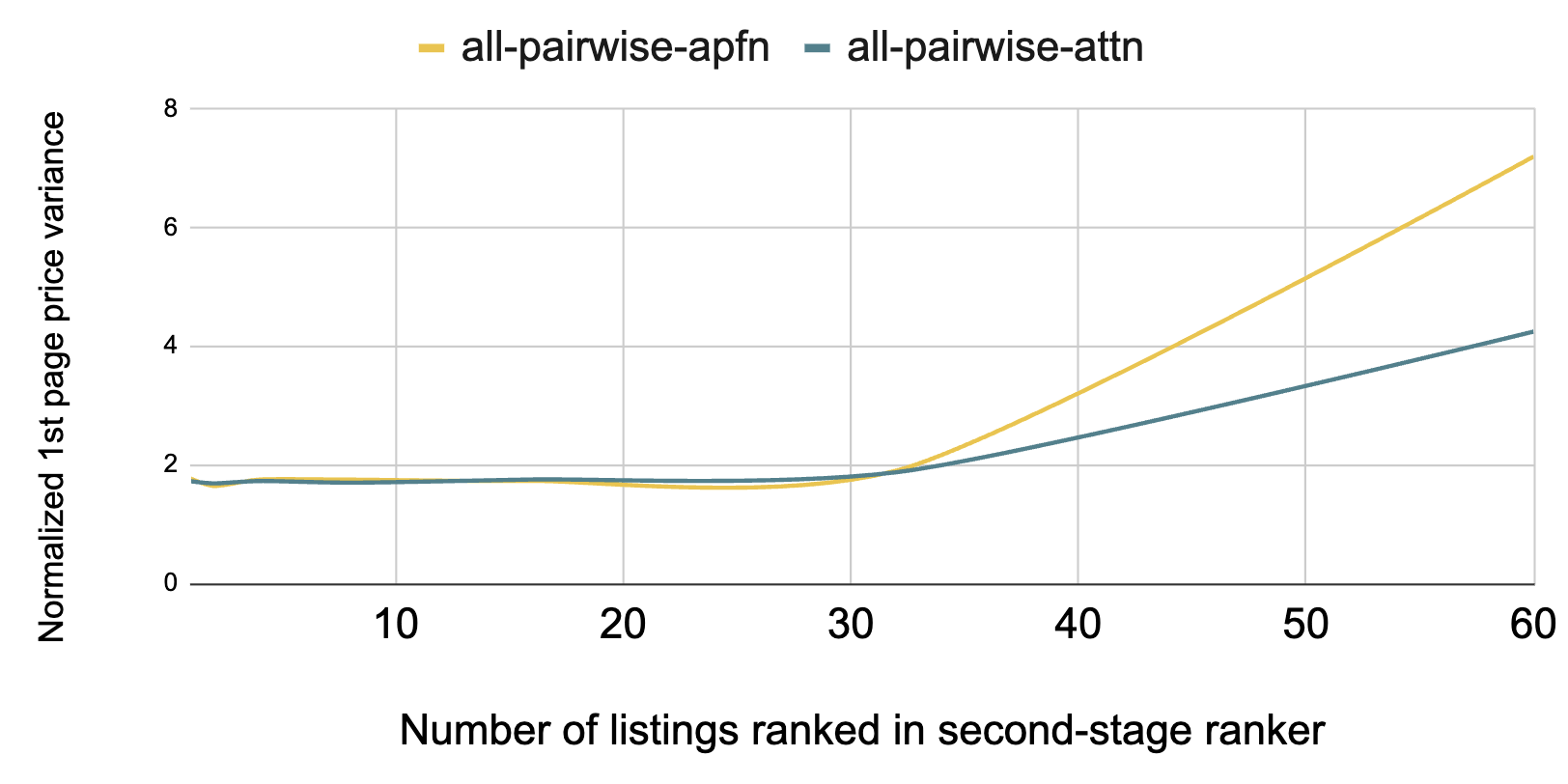}
\caption{\textmd{X-axis: Number of listings reranked. Y-axis: Normalized variance of listing prices on first page of search results.}}
\label{fig:diversity}
\end{figure}

Section~\ref{sec:allpair} explains how all-pairwise LTR improves ranking by computing listing similarities and optimizing for diversity. In our next experiment, we examine the impact of all-pairwise LTR on diversity, using the variance in listing prices on the first page of search results as a surrogate metric. Figure~\ref{fig:diversity} shows that diversity goes up as the all-pairwise LTR model is allowed to rerank more of the top listings.

While Figure ~\ref{fig:modelparamsndcg} clearly shows the superior performance of all-pairwise and true-pairwise LTR over pairwise LTR, in our fourth experiment we study a flip side---the uncertainty in model performance. Of the two varieties of uncertainty~\cite{NIPS2017_2650d608}---aleatoric and epistemic---we focus on measuring epistemic uncertainty, since aleatoric uncertainty is independent of the LTR frameworks. Table~\ref{tab:uncertainty} shows the range of NDCG per LTR framework based on changes in random seed for data shuffling and network initialization. The standard deviation in NDCG is around 2x to 6x higher compared to pairwise LTR. One possible reason for the higher epistemic uncertainty is that multivariate and bivariate models, with more inputs, can fit a broader range of parameters. However, averaging pairwise comparisons, as implemented in {\it true-pairwise-avg}, helps lower epistemic uncertainty.

\begin{table}[!h]
\centering
\begin{threeparttable} 
                 \begin{tabular}{c|c|c|c|c|c}
                   \toprule
                    $seed$ & \textbf{I} & \textbf{II} & \textbf{III} & \textbf{IV} & \textbf{V} \\
                    \midrule
                    1   & $0.6776$  & $0.6847$ & $0.6830$ & $0.6859$ & $0.6879$ \\
                    2   & $0.6777$  & $0.6838$ & $0.6833$ & $0.6856$ & $0.6874$ \\
                    3   & $0.6776$  & $0.6831$ & $0.6828$ & $0.6872$ & $0.6872$ \\
                    4   & $0.6772$  & $0.6833$ & $0.6827$ & $0.6867$ & $0.6852$ \\
                    5   & $0.6773$  & $0.6832$ & $0.6826$ & $0.6875$ & $0.6863$ \\
                    6   & $0.6779$  & $0.6841$ & $0.6826$ & $0.6864$ & $0.6877$ \\
                    7   & $0.6776$  & $0.6838$ & $0.6833$ & $0.6867$ & $0.6824$ \\
                    8   & $0.6781$  & $0.6836$ & $0.6831$ & $0.6865$ & $0.6862$ \\
                    \midrule
                    $stddev$ & $2.92\mathrm{E}{-4}$ &$5.29\mathrm{E}{-4}$ & $2.92\mathrm{E}{-4}$ & $6.23\mathrm{E}{-4}$ & $18.11\mathrm{E}{-4}$ \\
                   \bottomrule
            \end{tabular}
      \caption{\textmd{NDCG corresponding to variations of the models generated by different random seeds that control the training data shuffling and model parameter initialization. (I) pairwise LTR (II) true-pairwise-avg (III) true-pairwise-gbt (IV) all-pairwise-apfn (V) all-pairwise-attn. The final row computes the standard deviation of NDCG for the corresponding models.}}
      \label{tab:uncertainty}
\end{threeparttable}
\end{table}

\begin{figure}
\includegraphics[height=1.7in, width=2.55in]{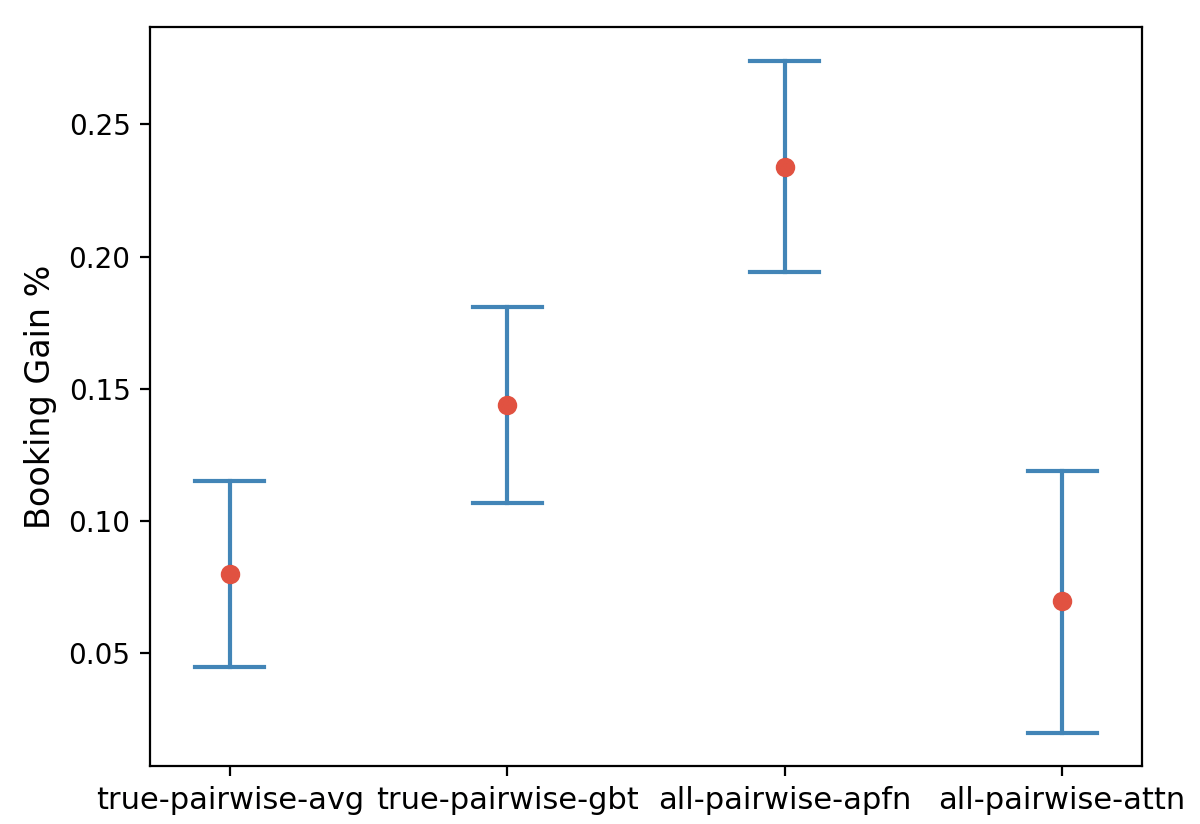}
\caption{\textmd{X-axis: Treatment models. Y-axis: \% gain in uncancelled bookings.}}
\label{fig:abtest}
\end{figure}

To measure how well offline performance generalizes to user impact, we perform online AB tests with pairwise LTR as control, against true-pairwise and all-pairwise LTR as treatments. We measure the percentage gain in uncanceled bookings over the control, which is shown in Figure~\ref{fig:abtest}. Given the statistical nature of the test, the $95\%$ confidence intervals are also shown.

The results in Figure~\ref{fig:abtest} show that offline NDCG does not generalize to online performance uniformly. By itself, NDCG provides an incomplete picture of model performance, so the final verdict must rely on key business metrics from online AB tests.

Our next experiment focuses on the multi-objective optimization discussed in Section~\ref{sec:multiobj}. We pick {\it all-pairwise-apn}, the most effective ranking model identified so far, and train it with the weighted loss from Equation~\ref{eq:weightedloss}. We compare it with Airbnb's production system, which used separate models to optimize bookings and trip quality. Results of the online AB test are highlighted in Table~\ref{tab:abtest}, which demonstrate the effectiveness of training all-pairwise LTR with the loss weighted by multi-objective.

\begin{table}[!h]
\centering
\begin{threeparttable} 
                 \begin{tabular}{l|c|c}
                   \toprule
                    Metric & Percentage Improvement & p-val \\
                    \midrule
                    Uncanceled Bookings & $0.61\%$  & $1\mathrm{E}{-5}$ \\
                    Uncanceled Nights Booked & $0.80\%$ & $1\mathrm{E}{-5}$ \\
                    Uncanceled Booking Value & $1.20\%$ & $1\mathrm{E}{-5}$ \\
                    Trip Rating Per Booking   & $0.03\%$  & $1\mathrm{E}{-5}$ \\
                    Bookings New Guest & $0.64\%$ & $0.03$ \\
                    Weighted Value  & $1.30\%$  & $1\mathrm{E}{-5}$ \\
                    \bottomrule
            \end{tabular}
      \caption{\textmd{Comparison of {\it all-pairwise-apn} trained with loss of Equation~\ref{eq:weightedloss} vs. active production ranker. }}
      \label{tab:abtest}
\end{threeparttable}
\end{table}

In our final experiment, we measure the improvement in stability discussed in Section~\ref{sec:stability}. To quantify stability, we introduce a small jitter to the map bounds associated with a search query. This introduces small randomized changes to the overall set of listings being ranked. We then count flips, where a top-ranked listing is replaced by one that was previously outside the top results. Comparing an all-pairwise LTR model without the residual architecture, unstable flips are reduced by $-75\%$ with the residual.

\begin{table}[!h]
\centering
\begin{threeparttable} 
                 \begin{tabular}{l|c|l|c}
                   \toprule
                       & Scalability & Accuracy & Total order \\
                    \midrule
                    Pairwise & $O(n)$  &  $\filledstar$& $\checkmark$ \\
                    True-pairwise & $O(n^2)$ & $\filledstar$$\filledstar$& $\times$ \\
                    True-pairwise** & $O(n^2)$ & $\filledstar$$\filledstar$&$ \checkmark$ \\
                    All-pairwise   & $O(n^2)$  & $\filledstar$$\filledstar$$\filledstar$& $\checkmark$ \\
                    \bottomrule
            \end{tabular}
      \caption{\textmd{Comparing LTR frameworks under the SAT theorem. True-pairwise** is modified by Equation~\ref{eq:tpbreak}, followed by generalized Bradley-Terry model.}}
      \label{tab:summary}
\end{threeparttable}
\end{table}

\section{Conclusion}
\begin{center}
``You can have two Big Things, but not three.''
\end{center}
\begin{flushright}
--- The Internet.
\end{flushright}

Of course, the internet is onto something here. But we found a working compromise: combine two things from two places. In our research, all-pairwise LTR as the second-stage ranker, combined with pairwise LTR in the first stage, emerged as the clear winner. Table~\ref{tab:summary} provides a summary. The all-pairwise LTR model from Section~\ref{sec:allpair}, which uses superiority-similarity features and trained with the multi-objective optimization from Section~\ref{sec:multiobj}, was successfully deployed to $100\%$ of Airbnb searchers in early 2025. Prior to the rollout, we conducted an in-depth review of key business metrics, system performance, and its impact on developer productivity. Two questions were identified that remain unanswered: (1) Why don’t all NDCG improvements lead to more bookings online---and what makes some types of NDCG gains matter more than others? (2) Are there techniques that allow all-pairwise LTR to scale further to handle a larger number of listings? Our current research is focused on these questions.